\documentclass{moriond}

\def\be{\begin{equation}}
\def\ee{\end{equation}}
\def\bea{\begin{eqnarray}}
\def\eea{\end{eqnarray}}

\begin{document}
\vspace*{4cm}
\title{Drell-Yan $p_{\bot}$ spectra from fixed-target to LHC energies \\ 
based on Parton Branching TMDs matched with NLO\\
\footnotesize{Contribution to the 2021 QCD session of the 55th Rencontres de Moriond}}

\author{ Aleksandra Lelek }

\address{Department of Physics, Particle Physics Group, Groenenborgerlaan 171,\\
2020 Antwerp, Belgium}

\maketitle\abstracts{
The theoretical description of Drell-Yan (DY) transverse momentum spectra over wide kinematic regions in energy, mass and transverse momentum 
requires not only fixed-order perturbative QCD calculations but also soft-gluon QCD resummations to all orders of perturbation theory.  The latter are 
traditionally accomplished either by Parton Showers (PS) with Monte Carlo event generators or by (different versions of)    
analytical procedures.  In this work we  focus on  
issues involved in the matching of  the fixed-order calculation and resummation, especially in the moderate to low mass and $p_{\bot}$ region. In particular 
we address  the DY region accessible 
at fixed target experiments. We present a Parton Branching (PB) formulation in which transverse momentum dependent (TMD) evolution is matched with 
MCatNLO calculations of NLO matrix elements. Using this formulation, we show a good theoretical description of DY data from experiments in very different 
kinematic ranges:  NuSea, R209, Phenix, LHC $8\;\textrm{TeV}$ and $13\;\textrm{TeV}$.}

\section{Introduction}

The Drell-Yan (DY) lepton-pair hadroproduction process is of crucial importance for particle physics. It is used as a standard candle for precision electroweak measurements at 
the LHC; it helps to understand the QCD evolution, resummation, factorization, both the collinear and transverse momentum dependent (TMD); it is used in parton distribution functions' (PDFs) extractions; at low masses and low energies, it allows one to access  information on partons'  intrinsic transverse momentum. However, the 
uniform theoretical description of DY data across a wide kinematic range in energy, mass and transverse momentum is highly non-trivial.

The baseline theoretical tool to obtain QCD predictions for production processes at high-energy hadron colliders is the 
collinear factorization formula \cite{Collins:1989gx}. Although this approach describes the structure of proton in the longitudinal direction only, it works perfectly well for sufficiently inclusive  hard processes characterized by a single mass scale. Still, for processes with more scales involved, such as the DY $p_{\bot}$ spectrum, also the transverse degrees of freedom have to be taken into account \cite{Angeles-Martinez:2015sea}. 
Different parts of the DY $p_{\bot}$ spectrum are driven by different physics: the high $p_{\bot}$ region (i.e. 
$p_{\bot}
{\raisebox{-.6ex}{\rlap{$\,\sim\,$}} \raisebox{.4ex}{$\,>\,$}}
%\sim 
Q$, where $Q$ is the invariant mass of the DY lepton pair) is expected to be described by fixed order QCD calculation within collinear factorization; in the low $p_{\bot}$ region ($p_{\bot}<<Q$), on the other hand, 
soft gluon radiation spoils the convergence of the fixed order  QCD calculation and  logarithms of $p_{\bot}/Q$ need to be resummed  to all orders in QCD running coupling $\alpha_s$. Formalisms which provide methods to perform soft gluon resummation are TMD factorization formulas (analytical Collins-Soper-Sterman (CSS) approach \cite{Collins:1984kg} or high energy ($k_{\bot}$) factorization \cite{Catani:1990xk,Catani:1990eg}) or Parton Shower (PS) procedures within Monte Carlo (MC) generators. The description of the intermediate $p_{\bot}$ region depends on the matching between the formalisms used in the low and high $p_{\bot}$. 

It was noted in the literature \cite{Bacchetta:2019tcu} that at fixed target experiments the  perturbative fixed-order calculations in collinear factorization are not able to describe the DY $p_{T}$ spectra for $p_{\bot}\sim Q $. In this work we examine this issue from the standpoint of  the Parton Branching (PB) approach \cite{Hautmann:2017xtx,Hautmann:2017fcj,Martinez:2018jxt} to TMD evolution.

\section{The Parton Branching method}
The PB method is a MC approach, based on the TMD factorization. It delivers the TMD PDFs (TMDs) which can be then used in TMD MC generators 
(like e.g. CASCADE \cite{Jung:2010si,Baranov:2021uol}) and collinear PDFs which can be used in standard collinear MC generators to obtain QCD collider predictions. 
The key element of the PB method is the TMD evolution equation 
which describes the change of the TMD $xf_a(x, k_{\bot},\mu)$ with the evolution scale $\mu$. Similarly to the 
showering \cite{Marchesini:1987cf}  version of the DGLAP equation \cite{Gribov:1972ri,Lipatov:1974qm,Altarelli:1977zs,Dokshitzer:1977sg},
it is  based on the unitarity picture where parton evolution is expressed in terms of real, resolvable branching probabilities, provided by the real emission DGLAP splitting functions and no-branching probabilities, represented by  Sudakov form factors.
Yet additionally to the longitudinal fraction $x$ of the protons' momentum, the PB evolution equation keeps track also of the parton's transverse momentum $k_{\bot}$. The starting distribution for the evolution includes the longitudinal momentum part and the intrinsic transverse momentum $k_{\bot 0}$ via a Gaussian factor,   and 
is fitted \cite{Martinez:2018jxt} to the HERA DIS data using \texttt{xFitter} \cite{Alekhin:2014irh}. Then, the transverse momentum is accumulated in each branching and at the end of the evolution it is a sum of $k_{\bot 0}$ and all the transverse momenta emitted in the evolution chain. Additionally, the approach includes soft gluon resummation by incorporating angular ordering (AO) \cite{Hautmann:2017fcj,Hautmann:2019biw}, similarly to \cite{Marchesini:1987cf}. The collinear PDFs $xf(x,\mu)$ are obtained from TMDs by integration over $k_{\bot}$. The obtained parton distributions are applicable in a wide kinematic range of $x$, $k_{\bot}$ and $\mu$, for all flavours. 

\section{DY $p_{\bot}$ spectra with PB TMDs + MCatNLO}
Once the TMDs are obtained, they can be used to derive QCD collider predictions. For that we use TMD MC generator CASCADE. As an example, let us concentrate on DY process. 
The idea is to promote the collinear factorization cross section formula written schematically  as  
\begin{equation}
\sigma = \sum_{ij}\int \textrm{d}x_1\textrm{d}x_2f_i(x_1, \mu^2)f_{j}(x_2, \mu^2)\hat{\sigma}_{ij}(x_1, x_2,  \mu^2, Q^2)
\end{equation} 
to a $k_{\bot}$-dependent formula 
\begin{equation}\sigma = \sum_{i j}\int \textrm{d}^2k_{\bot 1}\textrm{d}^2k_{\bot 2}\int \textrm{d}x_1\textrm{d}x_2f_i(x_1,k_{\bot 1}, \mu^2)f_{j}(x_2, k_{\bot 2}, \mu^2)\hat{\sigma}_{ij}(x_1, x_2,k_{\bot 1}, k_{\bot 2},  \mu^2, Q^2)
\end{equation}
where $\hat{\sigma}_{ij}$ is a partonic process, with $i$ and $j$ parton flavours.   
%\section{DY $p_{\bot}$ spectra with PB TMDs + MCatNLO}
In \cite{Martinez:2019mwt} the method  was developed to use PB TMDs with next-to-leading (NLO) ME within the  MADGRAPH5$\_$AMC@NLO (denoted here as MCatNLO) formalism \cite{Alwall:2014hca}.
If NLO cross section is to be combined with PS, care has to be taken about possible double counting. In order to do that, 
subtraction terms (for soft and collinear contribution) must be used. The exact form of the  
subtraction depends on the PS.
Because the PB TMDs have similar role to the PS, subtraction terms have  to be also applied to combine PB TMDs with NLO calculation.
Since PB uses AO which is similar to Herwig6 \cite{Corcella:2002jc}, MCatNLO is used with Herwig6 subtraction terms to combine  PB TMDs and NLO calculation. 
First, NLO subtracted ME is generated within MCatNLO using integrated PB TMD.
Then, an extra effort is needed to convert the collinear ME into a $k_{\bot}$-dependent one \cite{Martinez:2018jxt,Martinez:2019mwt}: the transverse momentum is added to 
the event record   within CASCADE according to the TMD and to conserve energy-momentum and keep the mass of the DY system unchanged, the longitudinal momentum 
fractions of the incoming partons have to be adjusted. For inclusive observables, like DY $p_{\bot}$ spectrum, 
the whole kinematics is included by using PB TMDs.

\begin{figure}
\begin{minipage}{0.33\linewidth}
\centerline{\includegraphics[width=0.86\linewidth]{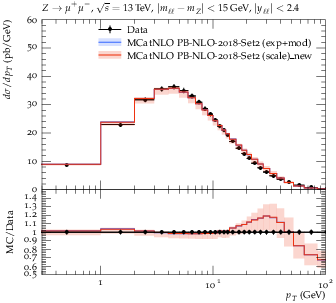}}
\end{minipage}
\hfill
\begin{minipage}{0.32\linewidth}
\centerline{\includegraphics[width=0.86\linewidth]{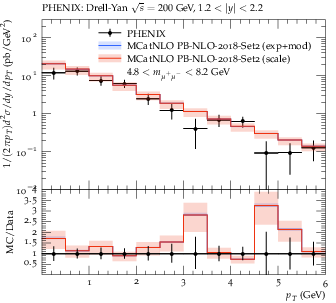}}
\end{minipage}
\hfill
\begin{minipage}{0.32\linewidth}
\centerline{\includegraphics[width=0.86\linewidth]{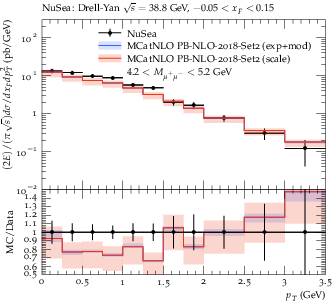}}
\end{minipage}
\caption[]{Predictions for DY $p_{\bot}$ spectra obtained with PB TMDs + MCatNLO compared with CMS (left), PHENIX (middle) and NuSea (right) data  \cite{Martinez:2020fzs}. }
\label{fig:predictions}
\end{figure}

The method was applied to describe the DY $p_{\bot}$ spectra coming from measurements at different center of mass energies and in different invariant mass ranges \cite{Martinez:2020fzs}: NuSea \cite{Webb:2003bj} ($\sqrt{s}=38.8\;\textrm{GeV}$), R209 \cite{Antreasyan:1981eg} ($\sqrt{s}=62\;\textrm{GeV}$), PHENIX \cite{Aidala:2018ajl} ($\sqrt{s}=200\;\textrm{GeV}$), LHC  \cite{Aad:2015auj,Sirunyan:2019bzr} (ATLAS $\sqrt{s}=8\;\textrm{TeV}$ and CMS $\sqrt{s}=13\;\textrm{TeV}$) and recently also Tevatron (D0 and CDF, $\sqrt{s}=1960\;\textrm{GeV}$). The results for NuSea, PHENIX and CMS are shown in Fig.\ref{fig:predictions}. Good theoretical description of the DY data coming from experiments in very different kinematic ranges was obtained with PB-NLO-HERAI+II-2018-set2 TMD PDF~\cite{Martinez:2018jxt} + MCatNLO. What is crucial, no additional adjusting of the method was performed to obtain predictions for low mass and low energy data compared to the procedure applied to LHC and Tevatron: the same PB TMD distribution was used for all the calculations and  no tuning of any parameters was involved.

\begin{figure}
\begin{minipage}{0.33\linewidth}
\centerline{\includegraphics[width=0.88\linewidth]{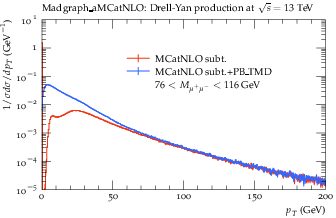}}
\end{minipage}
\hfill
\begin{minipage}{0.32\linewidth}
\centerline{\includegraphics[width=0.88\linewidth]{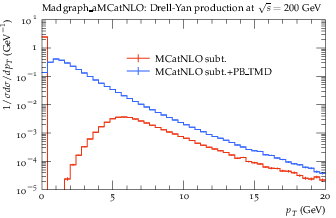}}
\end{minipage}
\hfill
\begin{minipage}{0.32\linewidth}
\centerline{\includegraphics[width=0.88\linewidth]{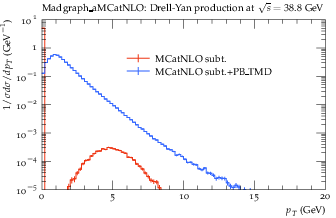}}
\end{minipage}
\caption[]{MCatNLO calculation with subtraction (red) and after including $k_{\bot}$ in the event record according to the PB TMD (blue) at the energies corresponding to CMS (left), PHENIX (middle) and NuSea (right) \cite{Martinez:2020fzs}. }
\label{fig:subtraction}
\end{figure}

In Fig.~\ref{fig:subtraction} the subtracted MCatNLO calculation (red) and the calculation after inclusion of $k_{\bot}$ in the event records according to the PB TMD using the method described above (blue) is shown for different center of mass energies. 
One can notice that 
at low DY mass and
low $\sqrt{s}$  even in the region of $p_{\bot} \sim Q$ the contribution of 
soft gluon emissions is essential to describe the
data. 
At larger masses and LHC energies the contribution from soft
gluons in the region of $p_{\bot} \sim Q$ is small and the spectrum driven by hard real emission.

\section{Conclusions}
It was noted in the literature that  perturbative fixed-order calculations in collinear factorization are not able to describe DY $p_{\bot}$ spectra at fixed target experiments in the region of $p_{\bot}\sim Q $.  Our observation is consistent with this remark: we notice that the contribution from soft gluons included in PB TMDs is essential to describe these data. 
The situation is different at LHC energies: here in the region of $p_{\bot}\sim Q$ the purely collinear NLO calculation gives a good result. 
The DY $p_{\bot}$ spectrum at low mass and low energy is sensitive to both fixed-order QCD and all-order soft gluon radiation, and the accuracy of the theoretical predictions depends on matching between those two. 
In the PB, the matching of PB TMDs and NLO calculation is done according to MCatNLO method: it is  
not an additive matching (as in CSS) but rather operatorial matching. TMD acts as an operator on subtracted NLO matrix elements, 
$\textrm{PB TMD} \otimes \left[ H^{(LO)} + \alpha_s\left( H^{(NLO)} - \textrm{PB TMD}(1)\otimes H^{(LO)}\right)\right]$. The
PB method
contains intrinsic, non-perturbative $k_{\bot 0}$ and well defined perturbative branching evolution, it includes
angular-ordered soft gluon radiation,  and it 
is matched through MCatNLO to NLO hard scattering. Thanks to all these elements, it
works in a wide DY kinematic range.

\section*{Acknowledgments}
The results presented in this article were obtained in collaboration with  A. Bermudez Martinez, P. Connor, D. Dominguez Damiani, L. I. Estevez Banos, F. Hautmann, H. Jung, J. Lidrych, M. Mendizabal Morentin, M. Schmitz, S. Taheri Monfared, Q. Wang, T. Wening, H. Yang and R. Zlebcik.
A. Lelek acknowledges funding by Research Foundation-Flanders (FWO) (application number: 1272421N).

\end{document}